\documentclass[a4paper,twoside]{article}
\newcommand{\affil}[1]{$^{\rm #1}$}

\usepackage[authoryear]{natbib}
\bibpunct{(}{)}{;}{a}{}{,}
\usepackage{graphicx}

\date{}

\begin{document}

\title{\large\bf\flushleft A Fuzzy Logic Based Algorithm For Finding Astronomical Objects in Wide-Angle Frames}

\author{\parbox{\textwidth}{\flushleft
\vspace{-0.5cm}
{\it Lior Shamir\affil{A,B}, Robert J. Nemiroff\affil{A}}\\
\vspace{0.4cm}
{\small \affil{A}\,Michigan Technological University, Department of Physics \\ 1400 Townsend Drive, Houghton, MI  49931}\\
{\small \affil{B}\,Email: lshamir@mtu.edu}}}

\vspace{0.4cm}

\twocolumn[
\begin{changemargin}{.8cm}{.5cm}
\begin{minipage}{.9\textwidth}
\vspace{-1cm}
\maketitle

\small{\bf Abstract:}
Accurate automatic identification of astronomical objects in an imperfect world of non-linear wide-angle optics, imperfect optics, inaccurately pointed telescopes, and defect-ridden cameras is not always a trivial first step. In the past few years, this problem has been exacerbated by the rise of digital imaging, providing vast digital streams of astronomical images and data.   In the modern age of increasing bandwidth, human identifications are many times impracticably slow.  In order to perform an automatic computer-based analysis of astronomical frames, a quick and accurate identification of astronomical objects is required. Such identification must follow a rigorous transformation from topocentric celestial coordinates into image coordinates on a CCD frame. This paper presents a fuzzy logic based algorithm that estimates needed coordinate transformations in a practical setting.  Using a training set of reference stars, the algorithm statically builds a fuzzy logic model. At runtime, the algorithm uses this model to associate stellar objects visible in the frames to known-catalogued objects, and generates files that contain photometry information of objects visible in the frame.  Use of this algorithm facilitates real-time monitoring of stars and bright transients, allowing identifications and alerts to be issued more reliably. The algorithm is being implemented by the Night Sky Live all-sky monitoring global network and has shown itself significantly more reliable than the previously used non-fuzzy logic algorithm.  


\medskip
\medskip
\end{minipage}
\end{changemargin}
]
\small

\section{Introduction}

Useful pipeline processing of astronomical images depends on accurate algorithmic decision making.  For previously identified objects, one of the first steps in computer-based analysis of astronomical pictures is an association of each object with a known catalog entry.  This necessary step enables such science as automatically detected transients and automated photometry of stars. Since computing the topocentric coordinates of a given known star at a given time is a simple and common task, transforming the celestial topocentric coordinates to image $(x,y)$ coordinates might provide  the expected location of any star in the frame. However, in many cases slight shifts in the orientation, inaccuracy of the optics or imperfections in the CCD can make this seemingly simple task formidable.

Fuzzy logic \citep{Zad65,Zad78,Zad83,Zad94} is an extension of Boolean logic that is useful for making complex decisions by computers.  While in Boolean logic an item has only two levels of memberships to a set (false or true; 0 or 1), fuzzy logic supports any value within the interval [0,1] as a level of membership of an item to a fuzzy set.  A fuzzy logic model consists of three different types of entities: fuzzy sets, fuzzy variables and fuzzy rules. The membership of a fuzzy variable in a fuzzy set is determined by a function that produces values within the interval [0,1]. These functions are called {\it membership functions}. Fuzzy variables are divided into two groups: {\it antecedent variables} that contain the input data of the fuzzy logic model, and {\it consequent variables} that contain the results calculated by the fuzzy logic model.

In this paper, we present a fuzzy logic based algorithm that transforms celestial coordinates into image coordinates for even complex combinations of wide-angle non-linear optical distortions, slight optical imperfections, and small unrecorded orientational perturbations.  In Section 2 we describe the fuzzy logic algorithm, and in Section 3 we present and discuss a practical implementation of the algorithm for automatic analysis of an operational astronomical project called ``Night Sky Live" \citep{Nem05}.

\section {Fuzzy Logic Based Coordinate Transformations}

The transformation of celestial topocentric spherical sky coordinates to local Cartesian image coordinates can be defined by a set of two functions.  Mathematical conversion from right ascension, declination, latitude, longitude, and local time to altitude and azimuth is straightforward.  Further transformation of azimuth and altitude sky coordinates to $(x,y)$ image (CCD) coordinates is frequently harder in practice when faced with practical inaccuracies of knowledge.  For practical application of fuzzy logic, this later transformation is broken up into two parts: 
\begin{equation}
(altitude, azimuth) \longmapsto x
\end{equation}
\begin{equation}
(altitude, azimuth) \longmapsto y
\end{equation}

On a CCD image, pixel locations can be specified in either Cartesian or polar coordinates.  Let $x_{\mathrm zen}$ be the $x$ coordinate (in pixels) of the zenith in the image, and $y_{\mathrm zen}$ be the $y$ coordinate of the zenith.  $x$ and $y$ coordinates of any given star visible in an astronomical image can be computed as follows: 
\begin{equation}
{x = x_{\mathrm zen} + \sin(angle) \cdot distance }
\end{equation}
\begin{equation}
{y = y_{\mathrm zen} + \cos(angle) \cdot distance }
\end{equation}
Where $angle$ is a polar azimuthal angle and $distance$ is a polar radial distance.

In order to use these equations it is necessary to compute a polar distance and angle for given objects. Given the observer's latitude and longitude, this can be done by converting the celestial coordinates (azimuth and altitude) of a given stellar object at a given time to the corresponding angle and distance in the image. Since the azimuth and altitude of any given bright star or planet at any given time can be easily computed, the only missing link here is the transformation of the altitude and azimuth to image coordinates, so the object can be found in the image. 

\subsection{Reference Stars}

Each of the two models is built based on manually identified reference stars. A reference star can be any star within an image that was correctly associated with the corresponding stellar object. Being familiar with the night sky, we can inspect the frame by eye and identify the stellar objects that appear in it. The image $(x,y)$ coordinates of the object can be taken from the peak of its point spread function (PSF). Each reference star provides a record with the following fields: azimuth, altitude, angle, and distance. 

Each identified star contributes an azimuth and altitude (by
basic astronomy) and also an angle and distance (by measurement from the
image). These provide the raw data for constructing a mapping between
the two, using the fuzzy logic model that will be described later in the paper. In order to obtain an accurate fuzzy logic model that calculates the angle, it is necessary to select reference stars that uniformly cover the entire image. This assures that the calculation of one value will depend on reference points that are relatively close to it. This is also true for the model that calculates the distance.

\subsection{Building the Fuzzy Logic Model}

In order to transform celestial coordinates into image coordinates, two different fuzzy logic models are being built based upon the two transformations:

\begin{equation}
f_{1}:azimuth \longmapsto angle  \newline
\label{f1}
\end{equation}
\begin{equation}
\label{f2}
f_{2}:altitude, azimuth \longmapsto distance 
\end{equation}
Here, {\it altitude}, {\it azimuth} and {\it angle} are angular measures, while {\it distance} is measured in pixels. Each transformation ($f_1$ and $f_2$) is computed by a different fuzzy logic model, thus one model calculates the angle and the other calculates the distance.  The fuzzy logic model $f_1$, which calculates the angle, is an approximation based on the assumption that the optical axis is perfectly aligned, so the angle is not depended on the altitude.

The fuzzy logic model $f_1$ has one antecedent (input) fuzzy variable and one consequent (output) fuzzy variable, while the fuzzy logic model $f_2$ has two antecedent variables ({\it altitude},{\it azimuth}) and one consequent fuzzy variable.

\subsection{Converting Azimuth to Polar Angle on the CCD ($f_1$)}

The first model ($f_1$) is built according to the reference stars such that each reference star adds to the model one fuzzy set and one fuzzy rule. Each fuzzy set is associated with a membership functions that is built in the form of a triangle \citep{Zad65}. Each of these membership functions reaches its maximum at the reference value, and intersects with the x-axis at the reference values of its neighboring reference stars. For instance, suppose we would like to build the fuzzy logic model with a data set that contains the following four reference stars: \newline

\def\0{\phantom{0}}
\halign{#\hfil& \quad#\hfil& \quad#\hfil& \quad#\hfil& \quad#\hfil& \quad#\hfil& \quad#\hfil& \quad#\hfil& \quad#\hfil& \quad#\hfil\cr
\bf azimuth& \bf altitude& \bf angle& \bf distance \cr
$0$& $\epsilon_0$& $\theta_0$& $R_0$\cr
$\alpha_1$& $\epsilon_1$& $\theta_1$& $R_1$\cr
$\alpha_2$& $\epsilon_2$& $\theta_2$& $R_2$\cr
$\alpha_3$& $\epsilon_3$& $\theta_3$& $R_3$\cr
}

\begin{figure}[h]
\begin{center}
\includegraphics[scale=1, angle=0]{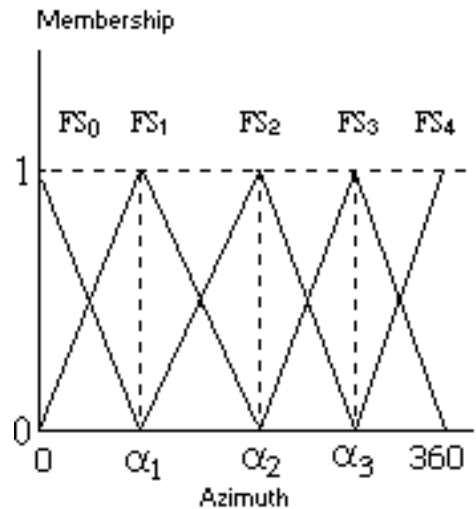}
\caption{The membership functions of the fuzzy sets ($FS_0$ to $FS_4$) created by the four reference values $(0,\theta_{0}),(\alpha_{1},\theta_{1}), (\alpha_{2},\theta_{2}), (\alpha_{3},\theta_{3})$ for $f_1(azimuth \longmapsto angle)$.}
\label{fuzzysets1}
\end{center}
\end{figure}

The first reference star maps azimuth $0^{o}$. Assuming $\alpha_1 < \alpha_2< \alpha_3$, the membership functions that will be added to the model are described in Figure ~\ref{fuzzysets1}.

The membership of the fuzzy sets is determined by the following membership functions: \newline \newline
$ F_0(x) = \cases{1 - {\frac {1} {\alpha_1}} \cdot $x$ & $0 \leq x \leq \alpha_1$ \cr
0 & $x<0$ or $x>\alpha_1$ \cr}
$
\newline \newline \newline
$ F_1(x) = \cases{{\frac {1} {\alpha_1}} \cdot $x$ & $0 \leq x < \alpha_1$ \cr
1 - {\frac {1} {\alpha_2-\alpha_1}} \cdot (x-\alpha_1) & $\alpha_1 \leq x \leq \alpha_2$ \cr
0 & $x<0$ or $x>\alpha_2$ \cr}
$
\newline \newline \newline 
$ F_2(x) = \cases{{\frac {1} {\alpha_2-\alpha_1}} \cdot (x-\alpha_1) & $\alpha_1 \leq x < \alpha_2$ \cr
1 - {\frac {1} {\alpha_3-\alpha_2}} \cdot (x-\alpha_2) & $\alpha_2 \leq x \leq \alpha_3$ \cr
0 & $x<\alpha_1$ or $x>\alpha_3$ \cr}
$
\newline \newline \newline 
$ F_3(x) = \cases{{\frac {1} {\alpha_3-\alpha_2}} \cdot (x-\alpha_2) & $\alpha_2 \leq x < \alpha_3$ \cr
1 - {\frac {1} {360-\alpha_3}} \cdot (x-\alpha_3) & $\alpha_3 \leq x \leq 360$ \cr
0 & $x<\alpha_2$ or $x>360$ \cr}
$
\newline \newline \newline 
$ F_4(x) = \cases{{\frac {1} {360-\alpha_3}} \cdot (x-\alpha_3) & $\alpha_3 \leq x \leq 360$ \cr
0 & $x<\alpha_3$ or $x>360$ \cr}
$
\newline \newline
$F_0$ to $F_4$ are the membership functions of the fuzzy sets $FS_0$ to $FS_4$ that were created according to the reference stars (the function $F_{m}$ is the membership function of the fuzzy set $FS_{m}$). The membership functions are built such that almost all azimuth values belong (with non-zero membership) to two fuzzy sets. Only the points of maximum ($0$,$\alpha_1$,$\alpha_2$,$\alpha_3$,$360$) have a non-zero membership to just one set. \newline \newline
The defuzzification method used in this model is {\it weighted average}, which is an efficient defuzzification method when the fuzzy logic model is built according to a set of signleton values \citep{Tak83,Tak85}. Since {\it weighted average} is used, the consequent part of each rule is a crisp value and not a fuzzy set. Therefore, the fuzzy rules that will be added to the model are: \newline
$FS_{0} \longmapsto \theta_0   $ \newline
$FS_{1} \longmapsto \theta_1   $ \newline
$FS_{2} \longmapsto \theta_2   $ \newline
$FS_{3} \longmapsto \theta_3   $ \newline
$FS_{4} \longmapsto 360   $ \newline
\newline
\newline
For instance, suppose that the first three reference stars have azimuths of $0{^o}$, $10{^o}$, and $20{^o}$, and their polar angles are $3{^o}$, $12{^o}$, and $22{^o}$ respectively, such that $\alpha_0=0{^o}$, $\theta_0=3{^o}$, $\alpha_1=10{^o}$, $\theta_1=12{^o}$, $\alpha_2=20{^o}$, and $\theta_2=22{^o}$. Each reference star adds one fuzzy set to the model such that the membership functions of the first two fuzzy sets $FS_0$, $FS_1$ are: \newline \newline
$ F_0(x) = \cases{1 - {\frac {1} {10}} \cdot x & $0 \leq x \leq 10$ \cr
0 & $x<0$ or $x>10$ \cr}
$
\newline \newline \newline 
$ F_1(x) = \cases{{\frac {1} {10}} \cdot x & $0 \leq x < 10$ \cr
1 - {\frac {1} {10}} \cdot (x-10) & $10 \leq x \leq 20$ \cr
0 & $x<0$ or $x>20$ \cr}
$
\newline \newline \newline 
Each reference star also adds one fuzzy rule such that the first three fuzzy rules are: \newline
$FS_{0} \longmapsto 3  $ \newline
$FS_{1} \longmapsto 12 $ \newline
$FS_{2} \longmapsto 22 $

Now suppose that we want to use this model in order to compute the polar angle (in the image) of a stellar object with an azimuth of $6{^o}$. The value $6$ has a level of membership of $1 - {\frac {1} {10}} \cdot 6 = 0.4 $ to the fuzzy set $FS_0$ and ${\frac {1} {10}} \cdot 6 = 0.6$ to the fuzzy set $FS_1$. Since the level of membership of 6 to all other fuzzy sets is 0, the only rules that will have any affect on the output of the computation are rules 0 and 1 above. Since the {\it weighted average} defuzzification method is performed, the output of the computation is ${\frac {0.4 \cdot 3 + 0.6 \cdot 12} {0.4+0.6}} = 8.4$.  This computation practically provides the same results as a simple linear interpolation of the value $6$ with the two closest neighboring reference stars.

\subsection{Converting Altitude and Azimuth to Radial Distance on the CCD ($f_2$)}

Unlike the simpler $f_1$ model used for transforming the azimuth to angle, the computation of the distance (in pixels) from $(x_{\mathrm zen},y_{\mathrm zen})$ should be computed based on {\it two} parameters, which are the altitude and the azimuth. Using both the altitude and azimuth allows the model to deal with asymmetric behavior of the optics as well as inaccurate orientational information.  In other words, using the assumption that the optics orientation is directly at the zenith and the distortion of the optics and hardware is completely symmetric, a reference point at a certain azimuth would allow calculating the distance of a point at the same altitude but at a different azimuth. However, this is not always the case. For instance, a stellar object with the azimuth of 0${^o}$ (north) and altitude of 30${^o}$ can be at distance of 150 pixels from $(x_{\mathrm zen},y_{\mathrm zen})$, while another stellar object at the same altitude (30$^o$) but at azimuth of 60${^o}$ will be at distance of 155 pixels from $(x_{\mathrm zen},y_{\mathrm zen})$. Moreover, $(x_{\mathrm zen},y_{\mathrm zen})$ does not necessarily appear in the center of the frame, and the frame is not necessarily centralized. Therefore, when computing the distance of a stellar object from $(x_{\mathrm zen},y_{\mathrm zen})$, it is required to take not only the altitude of the stellar object into considerations, but also the azimuth. In order to do that, the fuzzy logic model that calculates the distance is built according to reference stars that not only have different altitudes, but also different azimuths. 

We build the $f_2$ model that computes the distance using four different sets of reference stars such that each set contains reference stars that share approximately the same azimuth. For the sake of simplicity, the first set contains reference stars that are near azimuth 0$^o$, the second set contains reference stars near azimuth 90$^o$, and the other two sets contain stars near azimuth of 180$^o$ and 270$^o$ respectively. I.e., all reference stars used for this model should be fairly close to the azimuth of 0$^{o}$, 90${^o}$, 180$^{o}$ or 270$^{o}$. In order to use the four sets of reference stars, four new fuzzy sets are added to the model. Those fuzzy sets are ``North",``East", ``South", and ``West".  

\begin{figure}[h]
\begin{center}
\includegraphics[scale=1, angle=0]{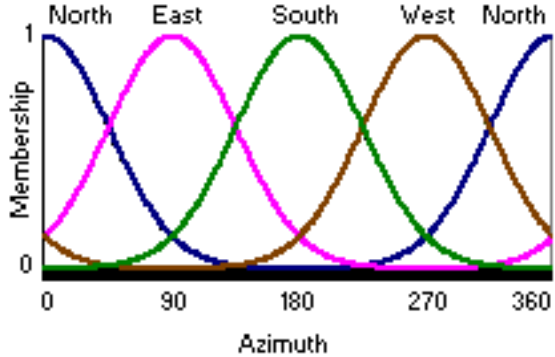}
\caption{The fuzzy sets of the four directions (North, South, East, West).}
\label{directions}
\end{center}
\end{figure}

Fuzzy logic can be viewed as a complex interpolation method.  In two dimensions, our fuzzy logic realization allows $f_2$ to perform the combined interpolation of linear and Gaussian functions in a straightforward fashion, and to alter the model easily when needed. Note that this is quite different than four separate linear and/or Gaussian interpolations for the four directions. The membership functions of the fuzzy sets are described in Figure~\ref{directions}.

The membership in the fuzzy sets is determined by the following Gaussian functions: \newline
\newline
North: \newline 
$ F_{\mathrm n}(x) = \cases
{ e^{-{\frac {x^2}{2\sigma^2}}} & $0 \leq x \leq 180$ \cr
  e^{-{\frac {(x-360)^2}{2\sigma^2}}} & $180 < x \leq 360$ \cr 
}
$
\newline
East: \newline
$ F_{\mathrm e}(x) = \cases
{ e^{-{\frac {(x-90)^2}{2\sigma^2}}} & $0 \leq x \leq 270$ \cr
  e^{-{\frac {(x-450)^2}{2\sigma^2}}} & $270 < x \leq 360$ \cr
}
$
\newline
South: \newline
$ F_{\mathrm s}(x) = e^{-{\frac {(x-180)^2}{2\sigma^2}}}$
\newline
West: \newline
$ F_{\mathrm w}(x) = \cases
{ e^{-{\frac {(x-270)^2}{2\sigma^2}}} & $90 < x \leq 360$ \cr
  e^{-{\frac {(x+90)^2}{2\sigma^2}}} & $0 \leq x \leq 90$ \cr
}
$
\newline \newline
Where $\sigma$ is set to $45$.

In this case, the level of membership for each of the four membership functions is always greater than zero, as opposed to the $f_1$ angle model where only two membership functions have a positive level of membership.
 
Building $f_2$ model can be demonstrated by an example: Suppose the model is built based on the following 6 reference stars listed in Table~\ref{sample_computation}.
\begin{table}[h]
\begin{center}
\caption{Reference stars}
\label{sample_computation}
\begin{tabular}{lccccc}
\hline star & azimuth & altitude & angle & distance\\
\hline 
$S_1$ & 0$^{o}$ & 68$^{o}$ & 2.4$^{o}$ & 215 pixels \\
$S_2$ & 359$^{o}$ & 62$^{o}$ & 1.2$^{o}$ & 224 pixels \\
$S_3$ & 1.5$^{o}$ & 72$^{o}$ & 3.4$^{o}$ & 206 pixels \\
$S_4$ & 90$^{o}$ & 66$^{o}$ & 94.2 & 180 pixels \\
$S_5$ & 91$^{o}$ & 62$^{o}$ & 95.6 & 188 pixels \\
$S_6$ & 91.6$^{o}$ & 70$^{o}$ & 95.9 & 172 pixels \\
\hline
\end{tabular}
\medskip\\
\end{center}
\end{table}

\begin{figure}[h]
\begin{center}
\includegraphics[scale=1, angle=0]{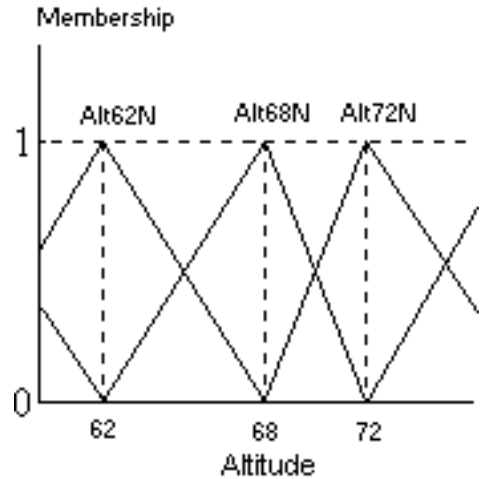}
\caption{The membership function of the fuzzy set {\it Alt68N}. The figure also include parts of the membership functions of {\it Alt62N} and {\it Alt72N} that were added by the two neighboring reference stars $S_2$ and $S_3$.}
\label{samp_memb_func}
\end{center}
\end{figure}

As with $f_1$, each reference star adds to the model one fuzzy set that has a triangle membership function. For instance, the membership function of the fuzzy set added by $S_1$ reaches its maximum of unity at 68, and intersects with the x-axis at the points of maximum of its neighboring reference stars. The neighboring stars are the two stars that their altitudes are closest to the altitude of $S_1$ (such that one is greater than 68$^{o}$ and one is smaller than 68$^{o}$), {\it and} have approximately the same azimuth as $S_1$. In this example, one neighboring star would be $S_2$ and the other would be $S_3$. All three stars share approximately the same azimuth (which is North). Therefore, the fuzzy set {\it Alt68N} that will be added by $S_1$ will have a triangle membership function that reaches its maximum at 68, and intersects with the x-axis at 62 and 72. This membership function is described in Figure~\ref{samp_memb_func}.

The membership function added by $S_1$ is: \newline
$ F_{Alt68N}(x) = \cases
{ {\frac {x-62} {68-62}} & $62 < x \leq 68$ \cr
  1- {\frac {x-68} {72-68}} & $68 < x \leq 72$ \cr
  0 & $x < 62$ or $x > 72$
}
$

Since the azimuth of $S_1$ is approximately north, the fuzzy rule added to the model by $S_1$ is: \newline
Alt68N $\land$ North $\longmapsto$ 215 \newline
This rule is more significant for stars that appear in the northern part of the sky (and in this case, also at an altitude of around $68{^o}$). The practical affect of this rule will be stronger as the coordinates are closer to the $68{^o}$ parallel. The fuzzy rules that will be added by the other reference stars are:\newline
Alt62N $\land$ North $\longmapsto$ 224 \newline
Alt72N $\land$ North $\longmapsto$ 206 \newline
Alt62E $\land$ North $\longmapsto$ 188 \newline
Alt66E $\land$ North $\longmapsto$ 180 \newline
Alt70E $\land$ North $\longmapsto$ 172 \newline

Now suppose that we want to use this model in order to compute the distance (in pixels) of a stellar object with an azimuth of $30{^o}$ and altitude of $64^{o}$. The value 64 have a membership value of ${\frac {64-62}{68-62}}={1 \over 3}$ to the fuzzy set {\it Alt68N} (that was added to the model by $S_1$), and $1 - {\frac {64-62}{68-62}}= {2 \over 3}$ to the fuzzy set {\it Alt62N} (that was added to the model by $S_2$). Also, this value would have a membership value of ${\frac {64-62}{66-62}}=0.5$ to the fuzzy set added by $S_4$, and a membership value of $1-{\frac {64-62}{66-62}}=0.5$ to the fuzzy set added by $S_5$. 
The membership value of the azimuth $30{^o}$ to the fuzzy set {\it North} is $e^{-{\frac {30^2}{2 \cdot 45^2}}}\simeq0.8$, and to the fuzzy set {\it East} it is $e^{-{\frac {(30-90)^2}{2 \cdot 45^2}}}\simeq0.412$. The membership values to the fuzzy sets {\it South} and {\it West} are very close to 0 in this case, and therefore have very little effect on the output value. For the sake of the simplicity of the example, these membership values are assumed to be exactly 0.  

The computation process is based on {\it product} inferencing and {\it weighted average} defuzzification. Therefore, the output value of the computation would be: \newline
${\frac {215 \cdot 0.8 \cdot 0.333 + 224 \cdot 0.8 \cdot 0.667 + 180 \cdot 0.412 \cdot 0.5 + 188 \cdot 0.412 \cdot 0.5}{0.8 \cdot 0.333 + 0.8 \cdot 0.667 + 0.412 \cdot 0.5 + 0.412 \cdot 0.5}}\simeq208.47$

\section{Example Application to Night Sky Live Data: Accuracy and Complexity}

The fuzzy logic based transformation formula has been tested and is in practical use with the Night Sky Live project (NSL) \citep{Nem05}.  Purposes of the global Night Sky Live (NSL) network of fisheye CONtinuous CAMeras (CONCAMs) include the ability to monitor and archive the entire bright night sky, track stellar variability, and search for transients.  Fully 2$\pi$ steradians -- half the sky -- are monitored passively by each CONCAM, without tracking.  Currently, the NSL project deploys nine CONtinuous CAMeras (CONCAMs) at many of the world's premier observatories.  When the Moon is down, CONCAMs take 180-second exposures every 236 seconds, and can detect stars down to visual magnitude 6.8 near the image center. 

The fuzzy logic based transformation formula is used by NSL for converting the celestial coordinates to image coordinates, so known catalogued stellar objects can be associated with PSFs that appear in the Night Sky Live frames. One simple task that is enabled by this transformation formula is the annotation of the all-sky images with the names of bright stars, constellations and planets. This task is mostly used for educational or ``cosmetic" purposes. A more important task is the automatic detection of non-catalogued objects. This task is required for automatic detection of meteors, comets, novae and supernovae, as well as other astronomical phenomena visible in the night sky. The following algorithm uses the transformation formula in order to associate PSFs in the image to stellar objects. \newline
\newline
1. function check\_stars(image, date\_time) \newline
2. image\_PSFs $\leftarrow$ GetPSFs(image) \newline
3. for each image\_PSF\_cords in image\_PSFs do \newline
4. begin \newline
5. $\indent$ min\_distance $\leftarrow \infty$ \newline
6. $\indent$ for each star in catalog do \newline
7. $\indent \indent$ star\_celestial\_cords $\leftarrow$ CelestialCoordinates(star, date\_time) \newline
8. $\indent \indent$ if InView(star\_celestial\_cords) then \newline
9.$\indent \indent \indent$ image\_cords $\leftarrow$ F(star\_celestial\_cords) \newline
10.$\indent \indent \indent$ if distance(image\_cords,image\_PSF\_cords) $<$ min\_distance then \newline
11.$\indent \indent \indent \indent$ min\_distance $\leftarrow$ distance(image\_cords, image\_PSF\_cords) \newline
12.$\indent \indent$ end if \newline
13.$\indent$ end for \newline
14.$\indent$ if min\_distance $< TOLERANCE$ then \newline
15.$\indent \indent$ associate(image\_PSF\_cords,star) \newline
16. end for \newline

The function {\it F} transforms celestial coordinates to image coordinates based on the fuzzy logic models described in this paper. The function {\it GetPSFs} returns a list of coordinates of the PSF peaks that appear in the picture. This function can be implemented by using some available algorithms for detection of sources from astronomical images such as {\it SExtractor} \citep{Ber95}. The function {\it InView} returns {\it true} if its argument coordinates are inside the relevant view of the optical device. In the inner loop the algorithm searches the catalog for a star that should appear closest to the center of the PSF. Since the hardware used for the Night Sky Live project currently cannot get deeper than magnitude 6.8, the catalog being used is a subset of Hipparcos catalog \citep{ESA97} that is sure to include objects this bright.  In line 14, the minimum distance found in the inner loop is compared with a constant value {\it TOLERANCE} that is a tolerance value. Only if the distance is smaller than {\it TOLERANCE} then  {\it image\_cords } and {\it image\_PSF\_cords } are considered as referring to the same star.

\begin{figure*}[h]
\includegraphics[scale=1, angle=0]{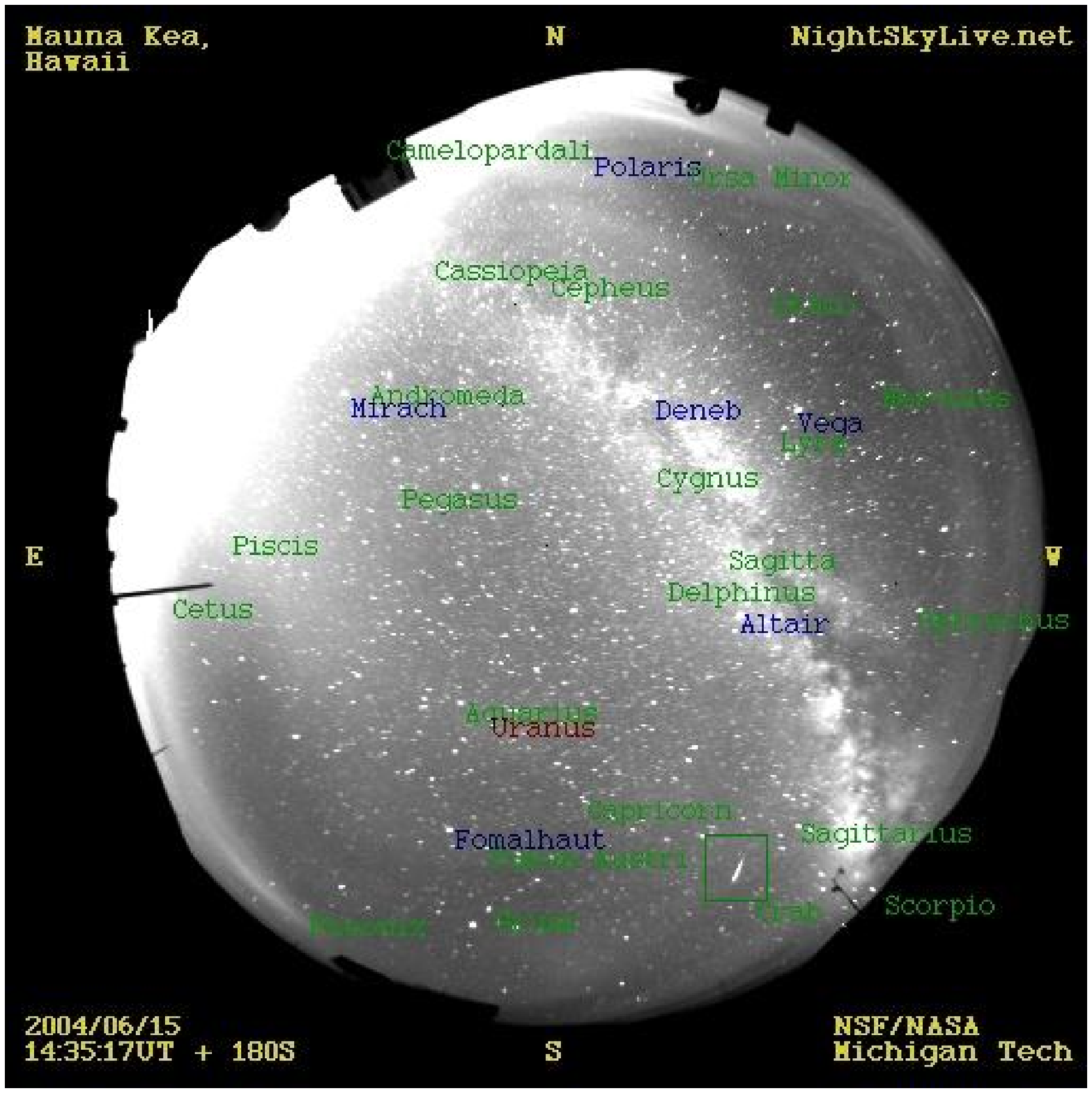}
\caption{A Night Sky Live picture with labeled stars, constellations, a planet and a meteor (boxed) processed using the fuzzy logic based transformation formula.}
\label{mauna_kea}
\end{figure*}

In Figure~\ref{mauna_kea} taken by the Mauna Kea CONCAM, the names of bright stars and constellations were annotated automatically using the above transformation.  The coordinates of one bright point (appears at the lower right of the frame) could not be associated with any catalogued object so it was automatically marked with a yellow square. This object is believed to be a meteor.  

The present fuzzy-logic algorithms allow practically 100 percent chance of accurate identification for NSL stars down to a magnitude of 5.6.  We are currently unaware of any exceptions.  A previously used NSL identification algorithm that employed a straightforward analytic transformation was only accurate to about magnitude 3.5, although that was somewhat dependent on the NSL station.  This dramatic improvement was the driving impetus for the creation of this paper.

A useful by-product of the newly accurate identifications is the automatic generation of photometry files. The ability to associate each PSF with a cataloged star allows the system to provide continuous monitoring of many bright stars. This information is provided in the form of text (XML tagged) files. Each frame produces one text file that lists all PSFs that were detected in the frame and the name and catalogue number of the stellar object associated with it. It also lists some additional data about each detected object such as the previously cataloged visual magnitude, spectral type, and celestial coordinates.  Identification allows other algorithms to process the frame and match each star with real time photometric data such as estimated counts of the background and the counts of the PSF. For each bright star, the average of the brightest 1, 5, 9, 16 and 25 pixels are listed. For dimmer stars, however, the realized PSF is much smaller so only the average of the top five pixels is listed. The on-line database allows browsing the records for past events. Since the photometry data itself is out of the scope of this paper, more information about it can be found at \citep{Nem05}.

The inverse computation of the presented transformation formula (converting angle and distance to altitude and azimuth) can be built in the same method described in this paper, with the exception of using the angle and distance for defining the membership functions, while the azimuth and altitude are used as the crisp output values of the fuzzy rules. This transformation formula is currently being used by NSL for computing 3D trajectories of meteors detected by the twin CONCAMs located at Mauna Kea and Haleakala.

\subsection{Computation Accuracy}

In order to check accuracy, the combined $f_1$ and $f_2$ models were used to compute the image coordinates of some 150 stars recorded by the NSL project \citep{Nem05}.  The NSL project uses fisheye images with an extreme viewable angle of 180${^o}$. For each pair of image $(x,y)$ coordinates, we calculated the Euclidean distance from the location calculated by the model and the location where the star appeared in the image. We took into considerations the average of the 150 Euclidean distances (the average error) and the worst case error, which is the longest distance among all computed 150 Euclidean distances.  Table~\ref{accuracy_level} below shows the accuracy levels according to the number of reference stars that were used in order to build both fuzzy logic models (the altitude $f_2$ model and the azimuth $f_1$ model).

\begin{table}[h]
\begin{center}
\caption{Accuracy Level of the Transformation as a Factor of the Number of Reference Stars}\label{accuracy_level}
\begin{tabular}{lcccc}
\hline Stars&Stars&Average Error& Maximum Error\\
in $f_2$&in $f_1$&(pixels) &(pixels)\\ 
\hline 
50& 40& 3.2& 4.8 \\
42& 40& 3.2& 4.8 \\
42& 30& 3.5& 5.5 \\
36& 30& 3.6& 5.7 \\
32& 20& 4.4& 6.2 \\
30& 20& 5.2& 8.0 \\
20& 20& 6.5& 10.2 \\
\hline
\end{tabular}
\medskip\\
\end{center}
\end{table}

\subsection{Computation Complexity}

The complexity of the computation is a function of the number of fuzzy rules and fuzzy sets in the model, which, in turn, is depended on the number of reference stars.  The accuracy needed to identify stars unambiguously is depended on the density of stars per pixel in the frame.  Clearly, the higher the average star density, the greater the accuracy needed to avoid confusion. 

Experiments suggest that accuracy within five pixels will allow automatic analysis of {\it Night Sky Live} frames.  As shown in Table~\ref{accuracy_level}, this can be achieved using a fuzzy logic model that is built with constant number of around 80 well spread out reference stars. Therefore, the final fuzzy logic model contains a constant number of around 80 rules and around 160 fuzzy sets, so the theoretical complexity of the computation is O(1). Practically, the CPU time that is required for the computation is negligible, and around 1000 coordinates are transformed in less than one second by a system equipped with an Intel Pentium III processor at 850 MHZ and 128MB of RAM.

\section{Conclusions}

Inaccuracies of optics and hardware lead to non-trivial transformation formulae that are sometimes required for the automatic analysis of the digital frames.  A fuzzy logic based method does not achieve an exact solution to the problem, but rather provides simple and maintainable transformation formulae with an accuracy that makes them more useful in practice than previously attempted straightforward analytic transformations (based on the linear transformation $R=k\epsilon$, where $R$ is the radial distance from $x_{\mathrm zen}$,$y_{\mathrm zen}$, $\epsilon$ is the altitude and $k$ is some constant).  When applied to the NSL project, the presented method is accurate enough to perform automatic processing of all-sky images in order to associate PSFs in the frames with their corresponding stellar objects nearly 100 percent of the time down to magnitude 5.6, and automatically detect non-catalogued bright objects. This technique demonstrates that it is possible to use fuzzy logic based algorithms to reduce the affect of minor defects and inaccuracies of the optical hardware and/or slightly inaccurate telescope orientation information. For NSL frames, only 80 reference stars are required to build the fuzzy logic model so that automatic identifications do not form a computational bottleneck.

\end{document}